\documentclass[preprint,showpacs,preprintnumbers,amsmath,amssymb,floatfix]{revtex4-1}

\usepackage{graphicx}
\usepackage{dcolumn}
\usepackage{bm}

\begin{document}

\title{Fission process of low excited nuclei with Langevin approach}

\author{Y.~Aritomo and S.~Chiba}

\affiliation{Research Laboratory for Nuclear Reactors, Tokyo Institute of Technology,
Ookayama, Meguro-ku, Tokyo, 152-8850, Japan}%




\begin{abstract}


Fragment mass distributions from the fission of U and Pu isotopes at low excitation energies are
studied using a dynamical model based on the fluctuation-dissipation theorem
formulated as Langevin equations.
The present calculations
reproduced the overall trend of the asymmetric mass distribution without parameter adjustment
 for the first time using the Langevin approach.
The Langevin trajectories show a complicated time evolution on the potential
surface, which causes the time delay of fission, showing that
dynamical treatment is vital.
It was found that the shell effect of the
potential energy landscape has a dominant role in determining the mass distribution,
although it is rather insensitive to the strength of dissipation.
Nevertheless, it is essential to include the effect of dissipation, since it
has a crucial role in giving ``fluctuation" to Langevin trajectories as well as for
explaining the multiplicities of pre-scission neutrons as the excitation energy increases.
Therefore, the present approach can serve as a basis for
more refined analysis.

\end{abstract}

\pacs{25.70.Jj, 25.85.w, 27.90.+b, 28.41.-i}



\maketitle



\section{Introduction}

The discovery of nuclear fission \cite{harn39,meit39} opened an important chapter
not only in the study of nuclear physics but also in the technology of energy supply.
Since the nuclear power plant accident at Fukushima in March 2011, further understanding of the
fission process has been required to quantitatively predict the amounts of heavy elements
and radioactive fission products remaining as ``debris" and the amount of melted spent nuclear fuel
still present in the remains of the power plant.  Moreover, such information
is also important for improving the safety of planned nuclear power plants, aiming at high burnup, world wide.
Therefore, further study of the nuclear fission process is necessary.

Shortly after the discovery of nuclear fission, it was interpreted in analogy with the fission of a charged
liquid drop; fission occurs as a result of competition between the disrupting effect of Coulomb repulsion
and the stabilizing effect of surface tension.
Bohr and Wheeler proposed this idea and invoked the liquid drop model to describe
the process \cite{bohr39}.  However, this concept could not explain asymmetric mass
splitting, which is the dominant mode of fission in nuclear fuel, namely,
U and Pu nuclei.

A fission model that uses the level densities of the nuclei at the ground
state and the saddle point was later developed, namely, the statistical model.
Using this model, it was possible to explain the qualitative features of the
mass-asymmetric splitting by introducing several phenomenological parameters \cite{fong,wilk}.
This model, however, did not include the dynamics of the fission process.
Moreover, pre-scission particles, particularly neutrons and $\gamma$-rays, may be emitted, which
alters the excitation energy and potential energy landscape of the fissioning system.
The large amount of experimental data accumulated in the 1980's
indicated that the pre-scission neutron
multiplicities from highly excited nuclei exceeded the values expected from the statistical model \cite{hils92}.
To explain this observation, the concept of dissipation,
which was proposed by Kramers in 1940 \cite{kram40}, was recalled.

By taking account of nuclear friction, which is the coupling between the fission
degree of freedom (collective motion) and other degrees of freedom such as nucleon
single-particle motion, it was
expected that a time delay exists that is necessary for a system to be deformed from a spherical shape to that of an elongated saddle, which provides time for
nucleons (mostly neutrons) to escape from the fissioning hot nuclei.
It was concluded that the pre-scission neutron
multiplicities in this picture significantly exceeded those predicted by the statistical model
and were in good agreement with observations \cite{hind89}.
On the other hand, mass-asymmetric fission, for example, by the thermal-neutron-induced fission of Th, U, and Pu nuclei,
 might be linked to the microscopic structure of fissioning nuclei
or fragments. However, the origin and mechanism of mass-asymmetric fission
have not yet been clarified.
Recently, the asymmetric fission of $^{180}$Hg was discovered following the electron
capture of $^{180}$Tl \cite{andr10}.
It was expected that symmetric fission would occur from the statistical model
picture owing to the strong shell effects of the half-magic nucleus $^{90}$Zr.
The fission paths, however, appeared to terminate before the system reached the
apparently dominant configuration of the two populating $^{90}$Zr nuclei.  The dynamics,
therefore, should play a key role in understanding fission.

To clarify the above contradiction and give a possibly unified picture of the fission process,
it is necessary to introduce a dynamical model of fission starting
from a nearly spherical shape and finishing at the scission region via the fission
saddle point.  Such a shape evolution proceeds in competition with pre-scission particle emissions; thus, a dynamical treatment is essential.
As such an approach, a method involving Langevin equations based on the fluctuation-dissipation theorem
has been applied to the nuclear fission process by several groups to describe the
process. This method takes account of
friction, inertia mass and multi-dimensional potential energy surfaces \cite{wada93,abe96,frob98,
vani99,chau02,karp03,schm03,ying07,mirf08,kilo08,sadh10,wang13}.
These past investigations focused on systems having high excitation energy.
The calculations resulted in a symmetric mass distribution of fission fragments (MDFF),
in good agreement with experimental data arising
from relatively high excitation energy.
The MDFF reflects the properties of the potential energy surface in the liquid drop model.
In contract, the dynamical calculation using Langevin equations has been seldom applied
to the fission process at low excitation energies \cite{asan04},
owing to difficulties in obtaining the shell correction energy of
configurations in the multi-dimensional space of collective coordinates,
as well as the huge computation time.
However, the computation time has recently been dramatically reduced
with the recent advances in computer technologies and the utilization of parallel
computing.
Moreover, we can calculate the shell correction
energy at each configuration using the two-center shell model.


In this paper, we present the possibility of dynamically calculating the fission process at a low
excitation energy using Langevin equations, taking into account the shell effects, pairing effects,
dissipation and fluctuation.
Using this model, we calculate the MDFFs of
$^{236}$U, $^{234}$U, and $^{240}$Pu at a low excitation energy and compare them with experimental data.
Using this approach, we can
investigate the fission mechanism, including the origin of mass-asymmetric fission.

The paper is organized as follows.
In Sec.~II, we detail the framework of the model.
In Sec.~III, we show the results for MDFF for
$^{236}$U, $^{234}$U, and $^{240}$Pu at the excitation energy $E^{*}=20$ MeV .
In Sec.~IV, we present a summary of this study and further discussion.

\section{Model}

We use the fluctuation-dissipation model and
employ Langevin equations \cite{arit04} to investigate the dynamics of the fission process.
The nuclear shape is defined by the two-center
parametrization \cite{maru72,sato78}, which has three deformation parameters, $z_0, \delta$, and $\alpha$ to serve as collective coordinates:
$z_{0}$ is the distance between two potential centers,
while
$\alpha=(A_{1}-A_{2})/(A_{1}+A_{2})$
is the mass asymmetry of the
two fragments, where
$A_{1}$ and $A_{2}$ denote the mass numbers of heavy and light fragments \cite{arit04}. The symbol
$\delta$ denotes the deformation of the fragments, and
is defined as
$\delta=3(R_\parallel-R_\perp)/(2R_\parallel+R_\perp)$, where $R_\parallel$ and $R_\perp$
are the half length of the axes
of an ellipse in the $z_0$ and $\rho$ directions of the cylindrical coordinate,
respectively, as shown in Fig.~1 in Ref. \cite{maru72}.
We assume in this work that each fragment has the same deformation.  This constraint
should be relaxed in the future work since the deformations of the heavy and light fragments in the fission of U region are known to be different from each other.
The deformation parameters $\delta$ and $\beta_{2}$ are related to each other as
\begin{equation}
\beta_{2}=\frac{\delta}{\sqrt{\frac{5}{16\pi}}(3-\delta)}.
\end{equation}
Notice that $\delta < 1.5$ since  $R_\parallel > 0$ and $R_\perp > 0 $.
In order to reduce the computational time, we employ the coordinate $z$ defined as
$z=z_{0}/(R_{CN}B)$, where $R_{CN}$ denotes the radius of a spherical compound nucleus
and $B$ is defined as $B=(3+\delta)/(3-2\delta)$.
We use the neck parameter $\epsilon=0.35$, which is recommended in Ref. \cite{sato78}
for the fission process.
The three collective coordinates may be abbreviated as $q$, $q=\{z,\delta,\alpha\}$.

For a given value of a temperature of a system, $T$,
the potential energy is defined as a sum of the liquid-drop (LD) part, a rotational energy and a microscopic (SH) part;
\begin{equation}
V(q,\ell,T)=V_{\rm LD}(q)+\frac{\hbar^{2}\ell(\ell+1)}{2I(q)}+V_{\rm SH}(q,T),
 \label{vt1}
\end{equation}
\begin{equation}
V_{\rm LD}(q)=E_{\rm S}(q)+E_{\rm C}(q),
\end{equation}

\begin{equation}
V_{\rm SH}(q,T)=E_{\rm shell}^{0}(q)\Phi (T),
\label{XevKK}
\end{equation}

\begin{equation}
\Phi (T)=\exp \left(-\frac{aT^{2}}{E_{\rm d}} \right).
\label{XevKK2}
\end{equation}
Here, $V_{\rm LD}$ is the potential energy calculated with the finite-range liquid drop model,
given as a sum of of the surface energy $E_{\rm S}$ \cite{krap79} and the Coulomb energy $E_{\rm C}$.
$V_{\rm SH}$ is the shell correction energy evaluated by Strutinski method from the single-particle levels of the two-center shell model.  The shell correction have a temperature dependence expressed by a factor $\Phi (T)$,
in which $E_{\rm d}$ is the shell damping energy chosen to be 20 MeV \cite{igna75} and
$a$ is the level density parameter.
At the zero temperature ($T=0$), the shell correction energy reduces to that of the two-center shell model values $E_{\rm shell}^{0}$.
The second term on the right hand side
of Eq. (\ref{vt1}) is the rotational energy
for an angular momentum $\ell$ \cite{arit04},
with a moment of inertia at $q$, $I(q)$.

\begin{figure}
\centerline{
\includegraphics[height=.80\textheight]{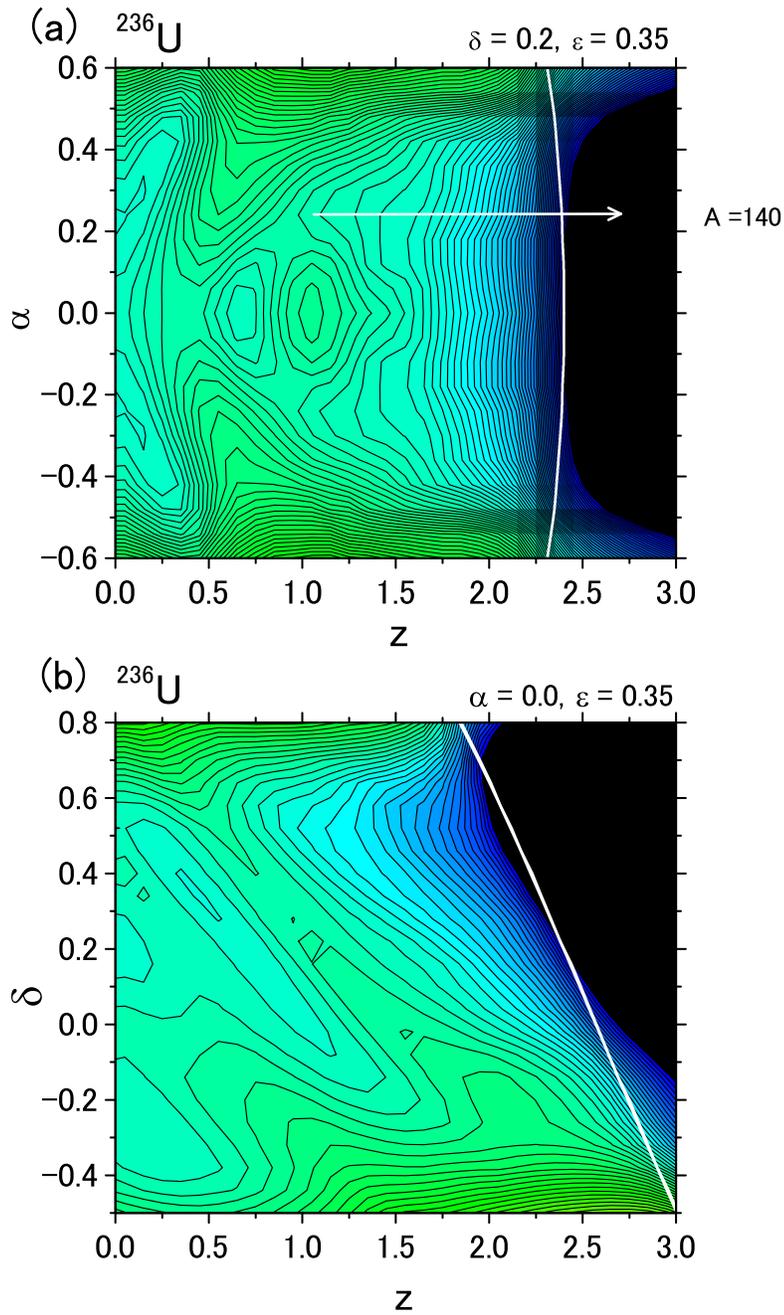}}
  \caption{(Color online) The potential energy surface $V=V_{\rm LD}+E_{\rm shell}^{0}$
with $\epsilon = 0.35$ in (a) the $z$-$\alpha$ space at $\delta = 0.2$ and
(b) the $z$-$\delta$ space at $\alpha = 0.0$ for $^{236}$U. The scission lines are denoted by the white lines.
The arrow indicates the fission valley that corresponds to $A = 140$. }
\label{fig_sza1}
\end{figure}

The multidimensional Langevin equations \cite{arit04} are given as
\begin{eqnarray}
&&\frac{dq_{i}}{dt}=\left(m^{-1}\right)_{ij}p_{j},\nonumber \\
&&\frac{dp_{i}}{dt}=-\frac{\partial V}{\partial q_{i}}
                 -\frac{1}{2}\frac{\partial}{\partial q_{i}}
                   \left(m^{-1}\right)_{jk}p_{j}p_{k} \nonumber \\
&&~~~~~~~~~~~~~~~~~~~~~-\gamma_{ij}\left(m^{-1}\right)_{jk}p_{k}\nonumber
                  +g_{ij}R_{j}(t),
\end{eqnarray}
where $i = \{z, \delta, \alpha\}$ and
$p_{i} = m_{ij} dq_{j}/dt$
is a momentum conjugate to coordinate $q_i$.
The summation is performed over repeated indices.
In the Langevin equation,
$m_{ij}$ and $\gamma_{ij}$ are the shape-dependent collective inertia and the
friction tensors, respectively.
The wall-and-window one-body dissipation
\cite{bloc78,nix84,feld87}is adopted for the friction tensor which can describe the
pre-scission neutron multiplicities and total kinetic energy of fragments simultaneously\cite{wada93}.
A hydrodynamical inertia tensor is adopted with the Werner-Wheeler approximation
for the velocity field \cite{davi76}.
The normalized random force $R_{i}(t)$ is assumed to be that of white noise, {\it i.e.},
$\langle R_{i}(t) \rangle$=0 and $\langle R_{i}(t_{1})R_{j}(t_{2})
\rangle = 2 \delta_{ij}\delta(t_{1}-t_{2})$.
The strength of the random force $g_{ij}$ is given by Einstein relation $\gamma_{ij}T=\sum_{k}
g_{ij}g_{jk}$.

The temperature $T$ is related with the intrinsic energy
of the composite system as $E_{\rm int}=aT^{2}$, where
$E_{\rm int}$ is calculated at each step of a trajectory calculation as
\begin{equation}
E_{\rm int}=E^{*}-\frac{1}{2}\left(m^{-1}\right)_{ij}p_{i}p_{j}-V(q,\ell,T=0).
\end{equation}
%

The fission events are determined in our model calculation by identifying the different
trajectories in the deformation space.
Fission from a compound nucleus is defined as the case that
a trajectory overcomes the scission point on the potential energy surface.
As an example, the potential $V_{\rm LD}+E_{\rm shell}^{0}$ with $\ell=0$ and $\epsilon = 0.35$
in the $z$-$\alpha$ space at $\delta = 0.2$ and in the $z$-$\delta$ space
at $\alpha = 0.0$ for $^{236}$U are shown in Figs.~\ref{fig_sza1}(a) and (b), respectively.
We define the scission point as the configuration in which the neck radius becomes zero, which is shown
by white lines in Fig.~\ref{fig_sza1}.
In Fig.~\ref{fig_sza1}(a), the arrow indicates the fission
valley that corresponds to $A = 140$.
We consider the neutron emission in our calculation.
However, we only take into account the decrease in the excitation energy of the system by neutron emission, not the change in the potential energy surface, as our first step.

\begin{figure}
\centerline{
\includegraphics[height=0.80\textheight]{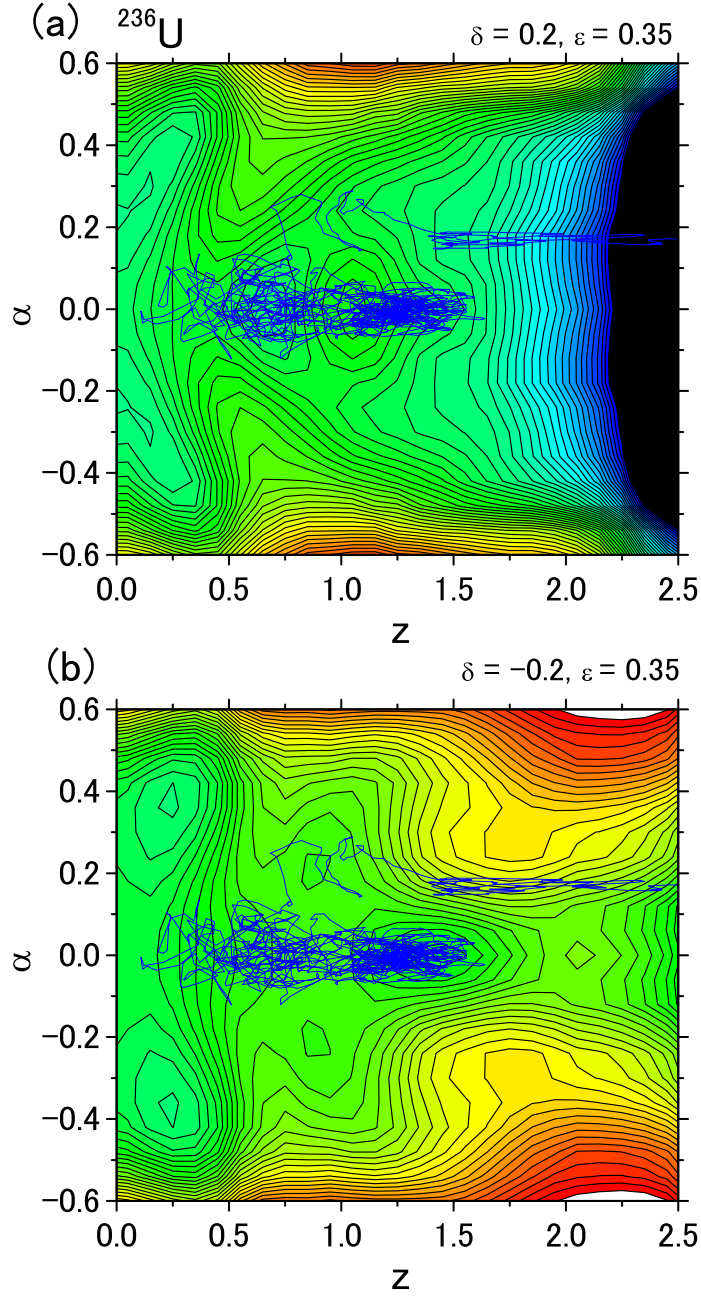}}
  \caption{(Color online) Sample trajectory of
$V_{\rm LD}+E^{0}_{\rm shell}$ for $^{236}$U projected onto the $z$-$\alpha$
plane at $\delta = 0.2$ (a) and $\delta = -0.2$ (b). The trajectory starts
at $z=0.65$, $\delta=0.2,$ and $\alpha=0.0$, at $E^{*}=20$ MeV, corresponding
to the second minimum
of the potential energy surface, to reduce the calculation time.}
\label{fig_traj1}
\end{figure}

\begin{figure}
\centerline{
\includegraphics[height=.40\textheight]{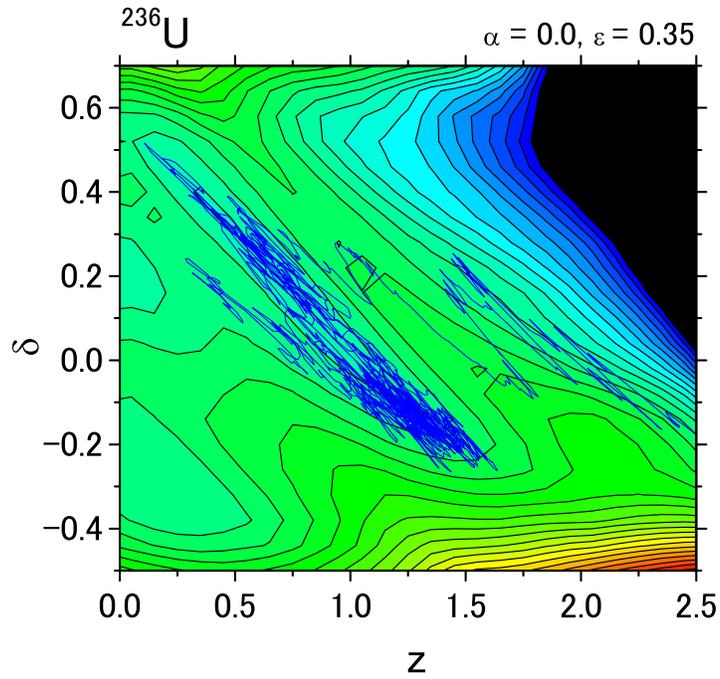}}
  \caption{(Color online)
The trajectory in Fig.~\ref{fig_traj1} is projected onto the $z$-$\delta$
 plane at $\alpha=0.0$. }
\label{fig_traj1c}
\end{figure}
\begin{figure}
\centerline{
\includegraphics[height=.40\textheight]{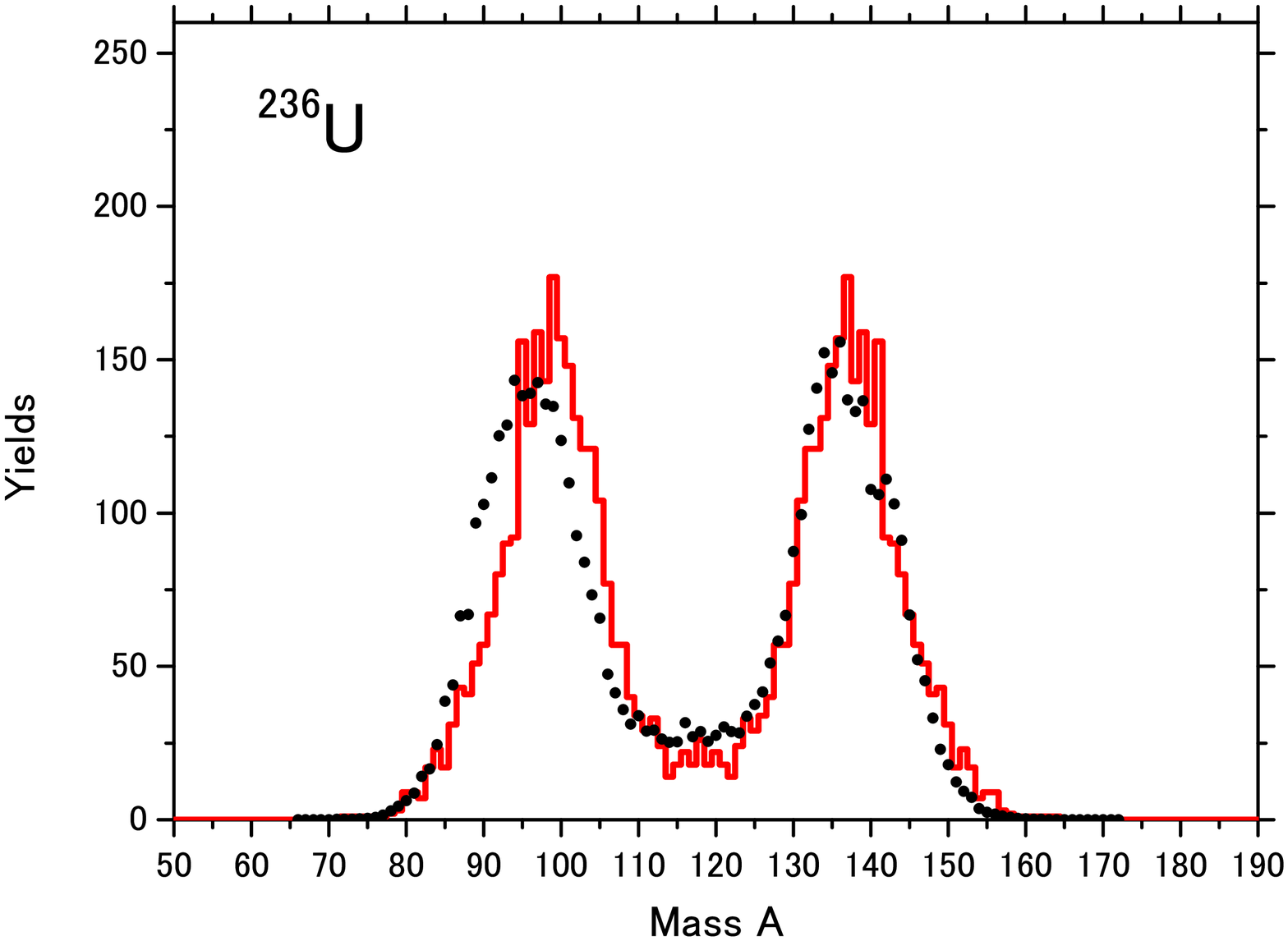}}
  \caption{(Color online) Mass distribution of fission fragments of $^{236}$U
  at $E^{*}= 20$ MeV.
  Calculation and experimental data are denoted by histogram and circles, respectively.}
\label{fig_sza3}
\end{figure}

\begin{figure}
\centerline{
\includegraphics[height=.40\textheight]{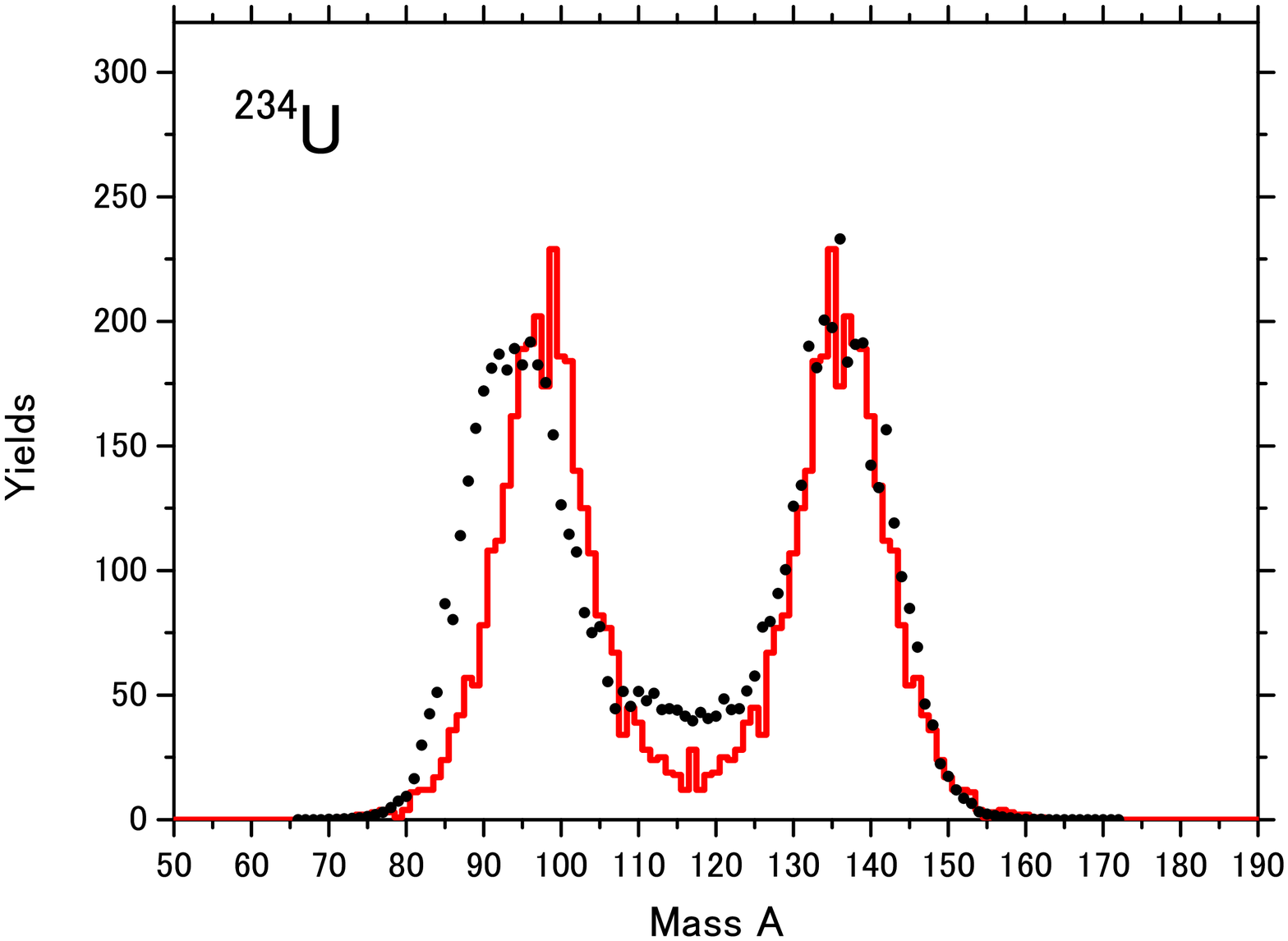}}
  \caption{(Color online)
Mass distribution of fission fragments of $^{234}$U
  at $E^{*}= 20$ MeV.
  Calculation and experimental data are denoted by histogram and circles, respectively. }
\label{fig_sza4}
\end{figure}

\section{Mass distribution of fission fragments}

Figure~\ref{fig_traj1} shows a sample trajectory of
$V_{\rm LD}+E^{0}_{\rm shell}$ for $^{236}$U that is projected onto the $z$-$\alpha$
plane at $\delta = 0.2$ (a) and $\delta = -0.2$ (b). The trajectory starts
at $z=0.65$, $\delta=0.2$ and $ \alpha=0.0$ at $E^{*}=20$ MeV, corresponding
to the second minimum
of the potential energy surface, to reduce the calculation time.
Indeed, the MDFFs thus obtained were equivalent to those starting from the ground state, namely, $z=0.0$, $\delta=0.2$, and $\alpha=0.0$.
The trajectory remains around the pocket located at $\{z, \delta, \alpha\} \sim \{1.35, -0.2, 0.0\}$
with thermal fluctuations.
Then, it escapes from the second minimum and moves along the valley corresponding to $A \sim 140$.
We project the trajectory in Fig.~\ref{fig_traj1} onto the $z$-$\delta$ plane at $\alpha=0.0$, as shown
in Fig.~\ref{fig_traj1c}.
The trajectory moves in the pocket around the second minimum and it remains at around $z \sim 1.35$ and $\delta
 \sim -0.2$ on this plane, which corresponds to the pocket in Fig.~\ref{fig_traj1}(b).
We will more precisely discuss the fission dynamics used to analyze the behaviors of trajectories in a forthcoming paper.

Figures~\ref{fig_sza3} and \ref{fig_sza4} show the calculated MDFFs for
$^{236}$U and $^{234}$U in the form of a histogram, together with the corresponding experimental data (dots) for neutron-induced
fission leading to the same compound nuclei at $E^{*}=20$ MeV, respectively.
The dots were taken from JENDL Fission Yield Data File  \cite{katakura} to represent the
experimental data concisely.
In the present calculation, we prepared 10,000 trajectories,
which is equivalent to the number of
trajectories of fission normalized by the total number of
fission events in the experimental data.
Here, we assume $\Phi (T)=1$ in Eq.~(\ref{XevKK}), which corresponds to the full shell
correction energy, to simulate the low excitation energy introduced by thermal neutrons. The effects of $\Phi (T)$ will be discussed later. For these nuclei,
the present approach yields results consistent with the measured data without any adjustment of the parameters in the Langevin calculation, showing the predictive power of the present model.
The widths and positions of the peaks are reproduced with high accuracy.
We consider that the trajectories move along the fission valley in Fig.~\ref{fig_sza1}(a),
as indicated by the arrow.
However, the peak of the light fragments in the calculation is
located at a position corresponding to a few more mass units.
This discrepancy is partly caused by the changes in the mass numbers of fissioning nuclei
upon the emission of neutrons from fragments, which is not included in our model.


\begin{figure}
\centerline{
\includegraphics[height=.40\textheight]{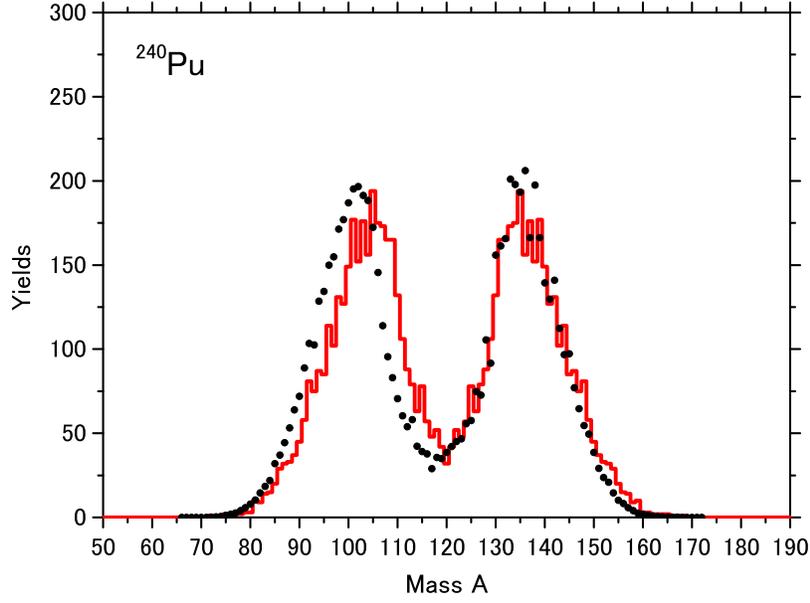}}
  \caption{(Color online) Mass distribution of fission fragments of $^{240}$Pu
  at $E^{*}= 20$ MeV.
  Calculation and experimental data are denoted by histogram and circles, respectively.}
\label{fig_sza77}
\end{figure}


We also calculate the MDFF of $^{240}$Pu at $E^{*}=20$ MeV, which is
shown in Fig.~\ref{fig_sza77} together with the corresponding experimental
data. The results are obtained using the same parameters as these in the calculations for $^{234}$U and $^{236}$U.
The results quantitatively agree with the experimental data, and the tendency of the difference
between the calculated results and experimental data is similar to the cases of $^{234}$U and $^{236}$U.


\begin{figure}
\centerline{
\includegraphics[height=.40\textheight]{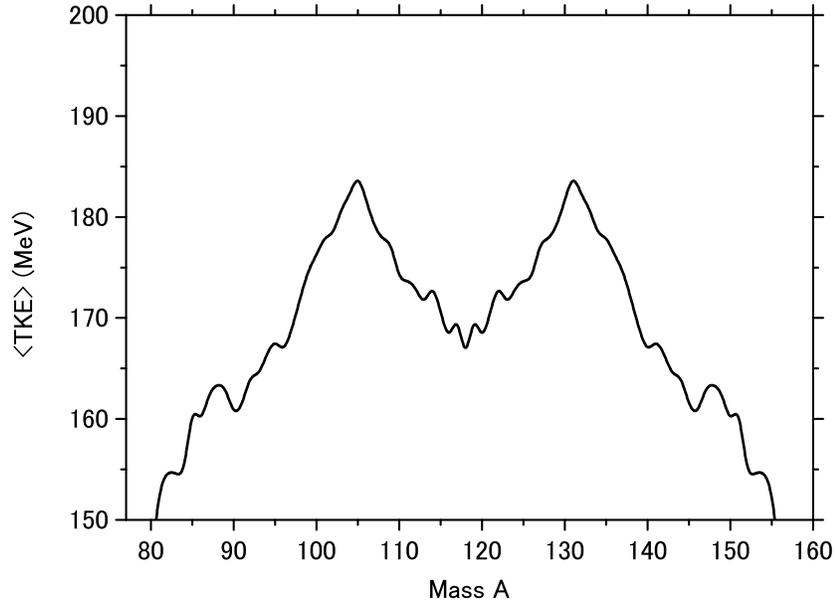}}
  \caption{Dependence of $\langle TKE \rangle$ on the mass number of the fission fragments of $^{236}$U
  at $E^{*}= 20$ MeV.}
\label{fig_tke}
\end{figure}

We calculate the average total kinetic energy of the fission fragments $\langle TKE \rangle$ of $^{236}$U at $E^{*}=20$ MeV.
We obtain $\langle TKE \rangle = 171.8$ MeV, which is consistent with the experimental data \cite{vand73}.
The dependence of $\langle TKE \rangle$ on the mass number of the fission fragments is shown in Fig.~\ref{fig_tke}.
The tendency observed is similar to the experimental data for the case of  $^{233}$U and $^{238}$U \cite{burn71}.


\begin{figure}
\centerline{
\includegraphics[height=.40\textheight]{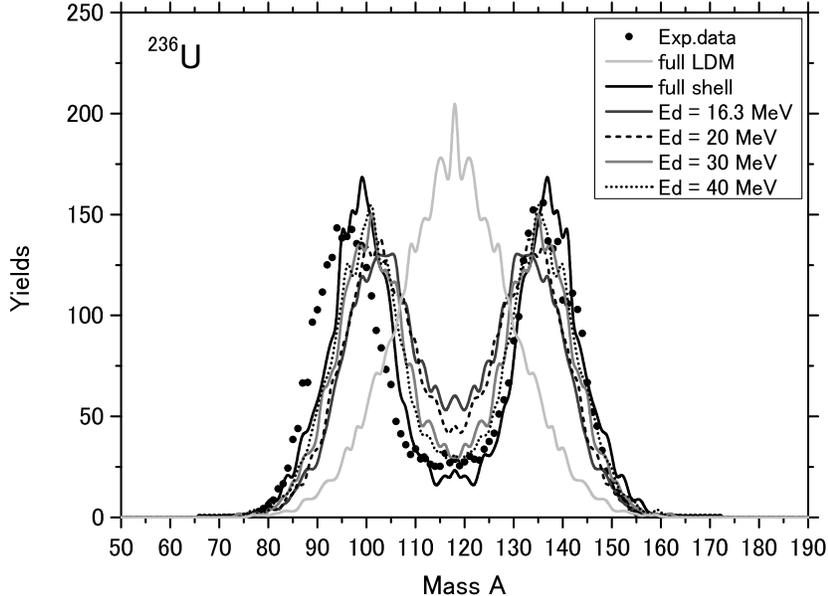}}
  \caption{Mass distribution of fission fragments of $^{236}$U
  at $E^{*}= 20$ MeV with the shell damping energy for $E_{d}= 16.3, 20, 30,$ and 40 MeV. The results with $V_{\rm LD}$(full LDM) and $V_{\rm LD}+E^{0}_{\rm shell}$ (full shell)
are denoted by the light gray and black lines, respectively. }
\label{fig_sza88}
\end{figure}

The shell correction energy depends on the excitation energy of the nucleus $E^{*}$, or the
nuclear temperature $T$.
We discuss the temperature dependence of the shell correction energy and how
the fission process and MDFF are affected. Considerable effort has been made to investigate the temperature
dependence of the level density parameter \cite{igna75,igna79}, which has been applied to the calculation of a statistical model for the fission process \cite{reis92}.
The temperature dependence of the potential energy surface
has been previously investigated  \cite{dieb81,loje85}.

Here, we assume that the temperature dependence of the shell correction energy is described by Eq.~(\ref{XevKK}) with the factor given by Eq.~(\ref{XevKK2}).
The shell damping energy of 20 MeV suggested by Ignatyuk et al. \cite{igna75} has not yet been confirmed \cite{jung98}.
Using several values of the shell damping energy, we investigate the effect of the MDFF on the shell dumping energy.
Figure~\ref{fig_sza88} shows the MDFF of $^{236}$U at $E^{*}$=20 MeV for $E_{d}= 16.3, 20, 30$, and 40 MeV.
When $E_{d}=16.3$ MeV, the effect of the shell correction energy is lower than when $E_{d}=40$ MeV
in this system.
Thus the MDFF with $E_{d}=16.3$ MeV produces a larger number of mass symmetric fission events than those in other cases owing to the effects of the potential energy surface of the liquid drop model.
However, the gross features of each case do not vary significantly.
Here, we plot the MDFF using $V_{\rm LD}$, which is denoted by the light gray line (full LDM). It shows mass-symmetric fission, which follows the structure of $V_{\rm LD}$.


\begin{figure}
\centerline{
\includegraphics[height=.40\textheight]{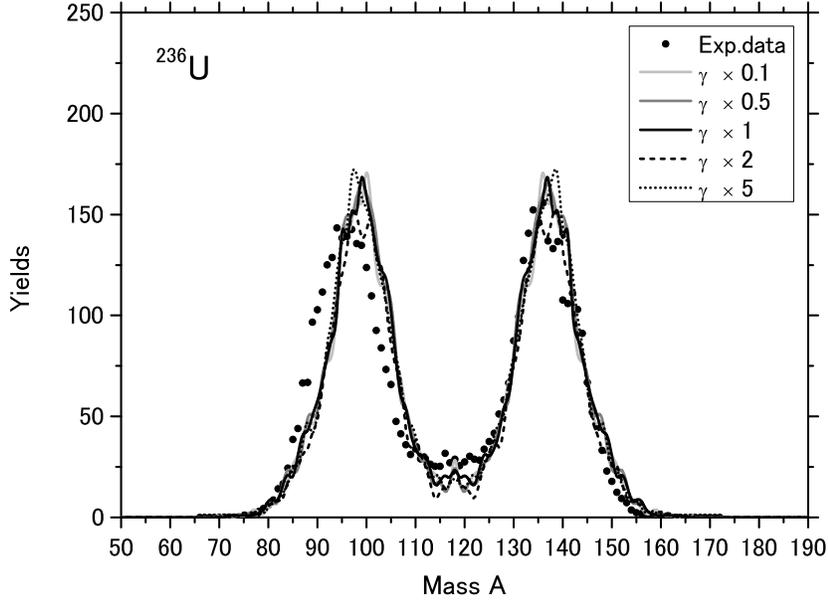}}
  \caption{Mass distribution of fission fragments of $^{236}$U
  at $E^{*}= 20$ MeV for each friction tensor.}
\label{fig_sza5}
\end{figure}

\begin{figure}
\centerline{
\includegraphics[height=.40\textheight]{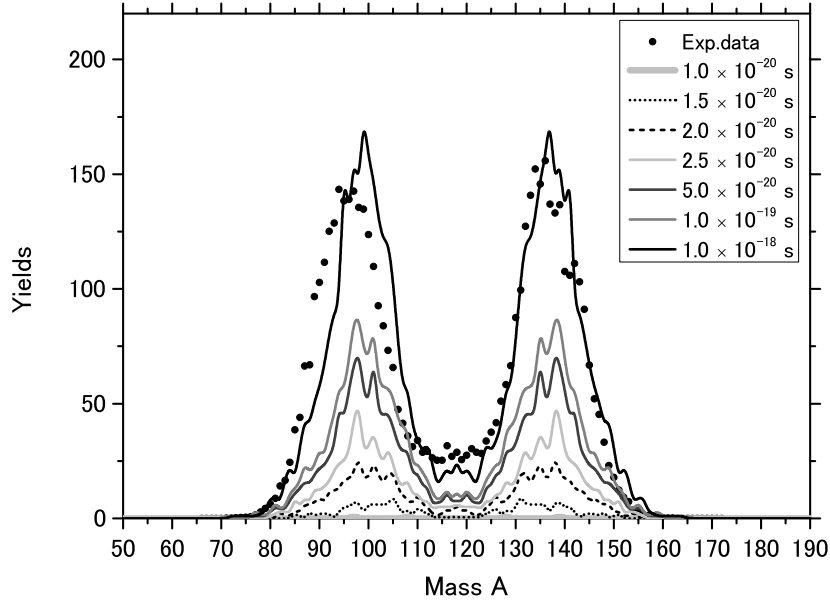}}
  \caption{Time evolution of mass distribution of fission fragments of $^{236}$U
  at $E^{*}= 20$ MeV. }
\label{fig_sza6}
\end{figure}

\begin{figure}
\centerline{
\includegraphics[height=.40\textheight]{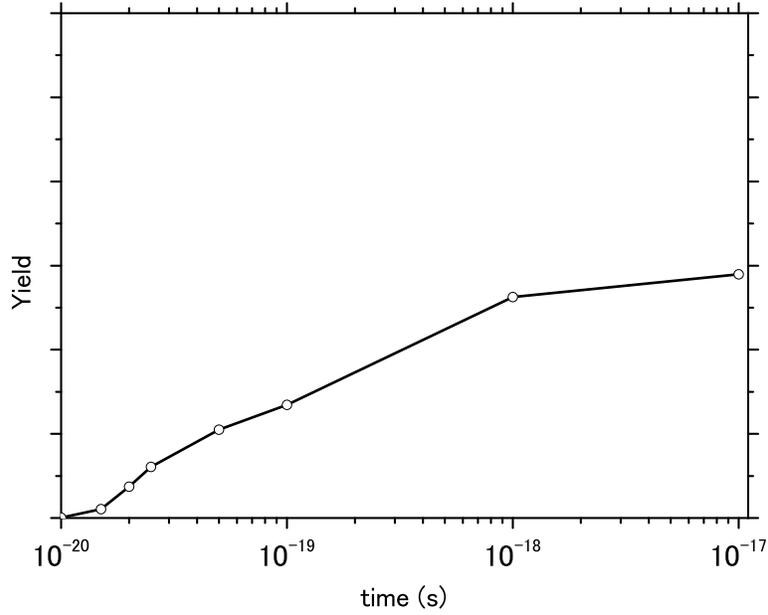}}
  \caption{
Time evolution of the number of fission events for $^{236}$U at $E^{*}= 20$ MeV. In the calculation,
we prepare 10,000 trajectories.}
\label{fig_time1}
\end{figure}

The MDFF is essentially governed by the dynamics of the trajectories in the potential
energy surface and is affected by the friction and inertia mass.
We investigate the MDFF of $^{236}$U at $E^{*}=20$ MeV by varying the strength of the
 friction tensor.
Figure~\ref{fig_sza5} shows the MDFF for the friction $\gamma$ multiplied
by factors of $0.1, 0.5, 1, 2,$ and 5.
Here, we assume $\Phi (T)=1$ in Eq.~(\ref{XevKK}).
The present results are rather insensitive to the strength of friction because the
excitation energy is low.
At a low excitation energy, there is little fluctuation of the trajectories, and the trajectories are mainly affected by the landscape of the
potential energy.


A major benefit of the dynamical calculation using Langevin equations is that we can
investigate the time scale of the fission process.
The time-dependent decay rate is governed by the nuclear collective dynamics, including fluctuation and
dissipation.
The study of the fission time scale is also important in
nuclear engineering since the emission of pre-scission neutrons, as a process competing with
fission,
alters the excitation energy
of the fissioning system; therefore it affects many phenomena such as the MDFF, the number of
prompt neutrons, their energy spectra, and the number of $\beta$-delayed neutrons.

The time evolution of the MDFF of $^{236}$U at $E^{*}=20$ MeV is shown in Fig.~\ref{fig_sza6}.
The trajectories do not reach the scission point until $t = 1.0 \times 10^{-20}$ s.
Then, the number of fission events increases with time.
Figure~\ref{fig_time1} shows the time evolution of the number of fission events for $^{236}$U at $E^{*}= 20$ MeV
with a logarithmic scale for time. Here, in the Langevin calculation, we prepare 10,000 trajectories.

We can see that almost all of the fission events occur dynamically until $t = 1 \times 10^{-18}$ s and that the number of fission events becomes saturated.
Actually, a complete Langevin description of the fission process must also consider the evaporation of light particles
and switches over to the statistical model description when the fission process reaches the stationary regime
\cite{mavl92}. This method can be used to treat all decay processes.
However, in the present study, we are interested in the MDFF of U.
To simplify the model, we do not switch the statistical model description, although we consider the
decrease in the excitation energy of the system during the dynamical calculation as being due to
the emission of neutrons.
This causes many Langevin trajectories to be trapped in the potential well for a long time because the excitation energies of such trajectories
are lower than the fission barrier height.
The fission lifetime is an important future subject of discussion.

\section{Summary}

In this study, we investigated the fission process at a low
excitation energy using Langevin equations.
We calculated the MDFF of $^{236}$U, $^{234}$U, and $^{240}$Pu at $E^{*}$=20 MeV, and the results
indicate mass-asymmetric fission, in good agreement with the experimental data,
without any parameter adjustment.
This is the first time that the MDFF has been obtained by a Langevin calculation while incorporating the shell effect and compared with experimental data.
In the present model, we used three collective variables to describe the shape of the nuclear fission.
We discussed the origin of the mass-asymmetric fission by analyzing sample trajectories.
This analysis allowed us to directly observe the time evolution of the dynamical process.

The calculated results depended slightly on the shell damping energy.  However, the dependence on the strength of the friction tensor was weak.  This does not mean that friction is
unimportant in the study of fission, since the variation of observables due to changes in the excitation energy is important, and, at higher energies, friction plays a very important role.  Therefore, inclusion of the effect of friction is important for a unified treatment of fission in a consistent manner and for applying the method to nuclei such as minor actinides, for which experimental data are extremely scarce.  At the same time, it was clarified that the
dynamical treatment is vital since the Langevin trajectories exhibit a rather complicated time evolution on the potential energy surface.
In particular, they spend a long time at the first and second potential minima, and
exhibit a ``fission time delay", during which competition with neutron evaporation may occur.

The reproduction of the experimental MDFF in this model can be considered as grounds for
supporting the investigation of fission dynamics at low excitation energies.
Furthermore, the generalized formula proposed in this model has the potential to simulate
any type of nuclear fission that may occur in the field of nuclear engineering.
Such simulation has become particularly necessary for applications 
to ensure the safe handling of nuclear waste and to improve the safety of planned nuclear power plants.

It should be noted that, a random walk method on the potential energy surface that incorporates the
shell correction energy was proposed and applied to the fission process at a low excitation
energy \cite{rand11}. This method also accurately reproduces the experimental mass yields
of $^{240}$Pu, $^{236}$U, and $^{234}$U.
Although this method appears to be a useful calculation tool, it can not be used to discuss the time scale of the fission process and its dynamics.

It is well known that the experimental MDFFs of $^{222,226}$Th have symmetric and triple-humped mass
distributions, respectively. Using our model, we attempted to calculate the MDFFs of both nuclei and obtained double-humped mass distributions for both nuclei, similarly to these for  $^{234,236}$U.
To describe these nuclei, we need a more realistic potential energy surface with more deformation variables.


In the future, we plan to improve the model to decrease the differences between the calculated MDFF and the experimental data.
We must increase the number of variables, at least by introducing independent deformation parameters
for each fragment. 
Moreover, the change in the potential energy owing to neutron emission from the fissioning system as well as from fission fragments should be included in the model.
Microscopic treatment of the
transport coefficients may also be important, particularly at low excitation energies, as carried out
in this study.
Since the computation time required to solve the Langevin equations will be increased by incorporating such improvements,
we must modify the computing algorithm to make it suitable for high-performance computers
and utilize parallel computing.
Still, the present approach can serve as a basis for such
more refined analysis aiming at providing a realistic description of the
entire process of fission, starting from compound nuclei at various excitation energies
and finishing at the population of fission products after $\beta$-decay, which is supported financially by MEXT through JST.



\section*{Acknowledgments}

Present study is the results of ``Comprehensive study of delayed-neutron yields for accurate
evaluation of kinetics of high-burn up reactors"
entrusted to Tokyo Institute of Technology by the Ministry of Education, Culture, Sports, Science and Technology of Japan (MEXT).
The authors are grateful to Prof. M.~Ohta, Prof. T.~Wada, Dr. A.~Iwamoto, Dr. K.~Nishio,
Dr.~A.V.~Karpov, Dr.~F.A.~Ivanyuk and Prof.~V.I.~Zagrebaev
for their helpful suggestions and valuable discussions.
Special thanks are deserved to Mr.~K.~Hanabusa (HPC Systems Inc.)  for his technical supports
to operate the high performance computer.

\vspace{10cm}
\newpage
\clearpage

\vspace{10cm}
\newpage
\clearpage

\end{document}